\documentclass[conference]{IEEEtran}

\usepackage{amsmath,amsthm,amssymb,paralist,subfigure,cite,graphicx,color,amsbsy,float}
\usepackage{pifont}
\usepackage{stmaryrd,mathrsfs,url}
\usepackage[table]{xcolor}
\usepackage{cuted,mathtools,lipsum}
\usepackage[ruled,vlined,linesnumbered]{algorithm2e}
\usepackage{soul}
\usepackage{microtype} 
\usepackage{graphicx}
\usepackage{hyperref}
\hypersetup{
	colorlinks=true,
	linkcolor=cyan,
	filecolor=mnodea,
	urlcolor=cyan,
	citecolor=lime,
}

\def\mb{\mathbf}

\def\mc{\mathcal}

\IEEEoverridecommandlockouts
\begin{document}
\title{\huge \bf
Distributed Observer Design for Tracking Platoon of Connected and Autonomous Vehicles  }

\author{\IEEEauthorblockN{ Mohammadreza Doostmohammadian}
\IEEEauthorblockA{
\textit{Faculty of Mechanical Engineering}, \\
\textit{Semnan University}, Iran, \\ \texttt{doost@semnan.ac.ir}}
\and
\IEEEauthorblockN{Hamid R. Rabiee}
\IEEEauthorblockA{\textit{Computer Engineering Department}, \\ \textit{ Sharif University of Technology},
	Iran, \\
	\texttt{rabiee@sharif.edu}}
}

\maketitle

\begin{abstract}
	Intelligent transportation systems (ITS) aim to advance innovative strategies relating to different modes of transport, traffic management, and autonomous vehicles. This paper studies the platoon of connected and autonomous vehicles (CAV) and proposes a distributed observer to track the state of the CAV dynamics. First, we model the CAV dynamics via an LTI interconnected system. Then, a consensus-based strategy is proposed to infer the state of the CAV dynamics based on local information exchange over the communication network of vehicles. A linear-matrix-inequality (LMI) technique is adopted for the block-diagonal observer gain design such that this gain is associated in a distributed way and locally to every vehicle. The distributed observer error dynamics is then shown to follow the structure of the Kronecker matrix product of the system dynamics and the adjacency matrix of the CAV network. The notions of survivable network design and redundant observer scheme are further discussed in the paper to address resilience to link and node failure. Finally, we verify our theoretical contributions via numerical simulations.        
		
\end{abstract}

\section{Introduction} \label{sec_intro}
Distributed algorithms have emerged as the new paradigm for computation, information processing, filtering, and control methods. This is because distributed (or decentralized) solutions advance traditional centralized schemes in terms of scalability, robustness to node failure, adaptability in large-scale applications, and enabling parallel processing. In a distributed system, the failure of one node does not necessarily compromise the entire system. This redundancy enhances reliability, making distributed algorithms more robust against failures \cite{tel2000introduction}. These algorithms are also more suitable for dynamic and heterogeneous setups, where agent/sensor nodes may fail or be added to the network. The rise of cloud and edge computing has further gained interest in distributed algorithms, as they align well with the architecture of these technologies \cite{duan2022distributed}.
Some recent applications include distributed fault detection \cite{ijc,chadli2017distributed}, distributed attack mitigation \cite{cao2023event,tnse_attack}, distributed machine learning and optimization \cite{khan2020optimization,ddsvm}, distributed routing via vehicular networks \cite{lin2020distributed,zhao2020routing}, distributed coverage control \cite{sayyaadi2010distributed}, and distributed resource allocation \cite{saidi2023task,cdc23}.  

Different distributed estimation and observer scenarios are proposed in the literature, see the survey in \cite{he2020distributed}. The existing methods include diffusion-based distributed estimation \cite{zamani2016iterative,zayyani2023adaptive}, same time-scale consensus plus innovation algorithms \cite{usman_cdc:10}, and delay-tolerant observer design \cite{delay_est,yoo2023adaptive}.
Cost-optimal design of sensor networks for distributed estimation is another interesting topic addressed in the literature \cite{spl18}. In the context of ITS, many works address different centralized and distributed control and estimation strategies for connected and autonomous vehicles (CAV). A hybrid model-data vehicle sensor and actuator fault detection and diagnosis system is considered in \cite{zabihi2024hybrid}. Observer-based control in a leader-follower vehicle platooning scenario is proposed in \cite{jiang2021observer}. String stability of
heterogeneous CAV Platoon under the multiple-predecessor following scenario is discussed in  \cite{abolfazli2022towards}. Originally introduced in the context of multi-agent systems, consensus algorithms have gained interest in CAV research. Consensus-based platooning under heterogeneous time-varying delays and switching topologies is considered in \cite{yu2022distributed}. Implementing model predictive control (MPC) offers a means to incorporate future predictions of vehicle behaviour within a distributed framework. Some MPC-based strategies are proposed in \cite{razzaghpour2021impact,ghanavati2023cloud}. Another approach involves the use of sliding mode control (SMC) techniques for robust control under disturbances. The work \cite{zhou2022decentralized} proposes a distributed SMC strategy for CAVs that handles uncertainties in vehicle dynamics, achieving improved synchronization and stability in platooning manoeuvres. Distributed reinforcement learning (RL) scenarios for merging in the mixed traffic environment of CAV and human-driven vehicles (HDV) are presented in \cite{shi2023deep}.  
Different distributed filtering techniques are also proposed for CAV tracking. For example \cite{shorinwa2020distributed} considers Kalman filter-like update to approximate the maximum a posteriori estimate without requiring the communication of measurements.  Due to its robustness in handling nonlinear dynamics and non-Gaussian noise, particle filtering has also been applied in CAV scenarios. The work \cite{ong2006decentralised} presents a decentralized particle filtering framework that allows flight vehicles to estimate in the presence of uncertainties. Distributed and delay-tolerant consensus-based filtering and tracking via the formation of unmanned aerial vehicles (UAVs) is considered in \cite{icrom22_track}.  

In this paper, we propose a distributed observer to track the global state of the CAV platoon. We first discuss the platooning model and its benefits in ITS setup. A linear dynamic system is then considered to model the CAV dynamics. The proposed distributed observer in this paper is defined over this linear CAV dynamics. The distributed observer/estimator is of a single time-scale, i.e., the estimation and vehicle dynamics follow the same time-scale. This is in contrast to double time-scale distributed observer \cite{he2020secure,battilotti2021stability,olfati2007distributed}, in which the estimation process takes place over a faster time-scale than the sampling time-scale of the vehicle's dynamics. This scheme, therefore, requires much faster and more expensive communication and computation devices than single time-scale methods. Moreover, the proposed distributed observer only requires global (or distributed) observability of the vehicle's dynamics. This is in contrast to many single time-scale methods that require local observability in the neighborhood of every vehicle \cite{das2015distributed,das2016consensus,samudrala2020distributed}, which mandates more communication links/channels and more network traffic. Our proposed distributed observer only needs strong-connectivity of the communication network of vehicles. This allows to design survivable networks and redundant observers resilient to node and link failure. We also verify this by numerical simulation in case we have arbitrary communication link failure or vehicle node failure.

The paper's organization is as follows. The dynamic model and the framework of the CAV platoon is described in Section~\ref{sec_fram}. The main observer design and estimation strategy with the gain design optimization method is discussed in Section~\ref{sec_dist}. Simulations to verify the theoretical contributions are given in Section~\ref{sec_sim}. Section~\ref{sec_con} provides the concluding remarks and some future research directions.

\section{The Framework of Vehicle Platooning Dynamics} \label{sec_fram}
\begin{figure} [hbpt!]
	\centering
	\includegraphics[width=3.5in]{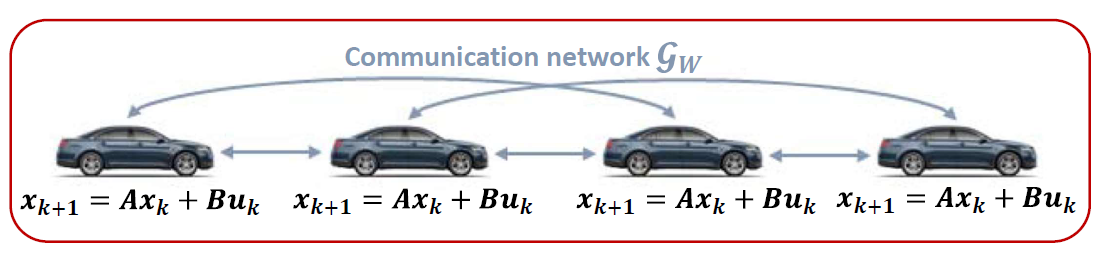}
	\caption{This figure shows the platoon of connected and autonomous vehicles communicating over a wireless network to share information. }
	\label{fig_platoon}
\end{figure}

Connected and automated vehicle (CAV) platooning is an advanced transportation strategy where multiple vehicles travel closely together in a coordinated formation, communicating with one another through vehicle-to-vehicle (V2V) and vehicle-to-infrastructure (V2I) technologies. An example of CAV platooning is given in Fig.~\ref{fig_platoon}.  
This system is designed to enhance traffic efficiency, improve safety, and reduce fuel consumption by optimizing vehicle spacing and speed. Vehicles are equipped with communication technologies that allow them to exchange state information data on position, speed, and acceleration with other vehicles. CAV utilize advanced sensors (like LiDAR, radar, and cameras) and control algorithms to operate with little or no human intervention. The combination of sensory data, communication devices, and control algorithms enables autonomous vehicles to adjust their speeds and positions relative to one another to maintain the desired spacing and ensure safety. The benefits of CAV platooning are reduced fuel consumption, reduced traffic congestion due to coordinated vehicle movements, and lower emissions due to improved fuel efficiency. Given these benefits, in this section, we define the dynamics, network, and sensor measurements for platoons of vehicles.

The CAV platoon is made of $n$ autonomous vehicles with position, velocity, and acceleration states denoted by $x^i = (p^i,v^i,a^i)^\top$. Then, its dynamics in continuous time follows as \cite{xiao2011practical},
\begin{align}  \label{eq_xdot}
	\dot{x}^i = A_cx^i + Bu^i = \left(
	\begin{array}{ccc}
		0 & 1 & 0\\
		0 & 0 & 1\\
		0 & 0 & -\frac{1}{\tau}	
	\end{array} \right)x^i + \left(
	\begin{array}{c}
		0 \\
		0 \\
		\frac{1}{\tau}	
	\end{array} \right)u^i,
\end{align} 
with $\tau$ as the time-constant denoting the driveline dynamics and $u^i$ as the input to vehicle $i$. Some other works in the literature, e.g. \cite{chehardoli2018third,bian2019reducing}, consider a constant velocity model, which is more restrictive than the model in this paper. Using  a sampling mechanism with sampling interval $T$, the discretized dynamics of every vehicle $i$ is defined as
\begin{align}  \nonumber
	x^i_{k+1} &= Ax^i_k + Bu^i_k \\ \label{eq_xdot_d}
	&= \left( 
	\begin{array}{ccc}
		1 & T & 0\\
		0 & 1 & T\\
		0 & 0 & 1-\frac{T}{\tau}	
	\end{array} \right)x^i_k + \left(
	\begin{array}{c}
		0 \\
		0 \\
		\frac{1}{\tau}	
	\end{array} \right)u^i_k,
\end{align}
with $k$ as the discrete time-instant.
The platooning distance between vehicle $i$ and its neighboring vehicle $i-j$ is considered as
 \begin{align}  \label{eq_di}
 	d^{i,i-j}_k = \sum_{l=i}^{i-j+1} h_l v^{l-1} + jD,
 \end{align} 
with $D$ as the static gap between two consecutive vehicles, and $h_l$ as the time-headway of the vehicle $l$ in the platoon. This paper considers a homogeneous CAV platoon, i.e., all vehicles are similar and have the same dynamics, speed characteristics, and control design. Therefore, one can replace $h_l=h$ in Eq.~\eqref{eq_di}. The dynamics is associated with some noisy sensor measurements on the position of the vehicles, i.e., the output measurement is defined as,
\begin{align}  \label{eq_yi}
	y^i_k = C_i x^i_k+r_k=(1,0,0)x^i_k,
\end{align}
with $r_k$ as the output noise at time $k$. The global sensor output is $y_k = Cx_k$ with $C = (C_1,\dots,C_n)^\top$.    
Define the position of each vehicle with its predecessors over the CAV platoon as
 \begin{align}  \label{eq_pi}
	p^{i}_k = p^{i-1}_k - d^{i,i-1}_k,
\end{align}
with $d_{i,i-1}$ as in Eq.~\eqref{eq_di}. Different control strategies are proposed in the literature to determine $u_i$ at each vehicle $i$ to ensure the string stability of the vehicles\footnote{The main objective of this paper is on distributed observer design to track the state of CAV platoon and the control input design is not the focus of this paper. Therefore, in this section, we only review the existing control design strategies in the literature.}. For example, \cite{jiang2021observer} considers a linear feedback controller which only needs local information exchange over the CAV network. Some other control strategies include: PID controllers to guarantee stable platooning \cite{chehardoli2018third,bian2019reducing}, model predictive control (MPC)  \cite{razzaghpour2021impact,ghanavati2023cloud}, and event-trigerred control via barrier functions \cite{sabouni2024optimal}. Some other works assume nearly-constant-acceleration (NCA) model and consider the input to be unknown modelled by random variations, see \cite{icrom22_track} for example.  

The network over which the vehicles communicate relevant information is denoted by a graph $\mc{G}_W$ with the adjacency matrix $W$ denoting the weight associated with the information received from neighboring vehicles. In other words, every vehicle sets a weight factor on the state and estimation information of its neighbors. In this direction, a matrix
$W = \{w_{ij}\}$ denotes the adjacency  matrix of $\mc{G}_W$. Following the consensus literature \cite{olfatisaberfaxmurray07}, this matrix satisfies stochastic property, i.e.,
\begin{align}  \label{eq_wi}
  W \mb{1}_n = \mb{1}_n \longleftrightarrow	\sum_{j=1}^n w_{ij} = 1,~\forall i=1,\dots,n
 \end{align}
with $\mb{1}_n$ as the column vector of all ones of size $n$. This consensus matrix guarantees agreement on the estimation of the CAV platoon at all vehicle nodes.

The overall dynamics of the CAV is considered as a block diagonal system matrix with each diagonal block defining the dynamics of one vehicle. This follows the assumption that the vehicle platoon is homogeneous, implying the same dynamics for all vehicle nodes. This document aims to design a distributed observer to track the state of this large-scale system as a dynamic model for the vehicle platoon. The structure of such an observer needs to follow the communication network $\mc{G}_W$ of vehicles while addressing observability concerns in a distributed setup. This is addressed in detail in the next section.  

\section{The Proposed Distributed Observer} \label{sec_dist}
There are two main schemes for distributed estimation and observer design: double time-scale method \cite{he2020secure,battilotti2021stability,olfati2007distributed} and single time-scale method \cite{das2015distributed,das2016consensus,samudrala2020distributed}. In the first case, sensors perform many steps of consensus-based data-fusion between every two consecutive samples of CAV dynamics. This requires much faster devices to perform the communication and consensus filtering in between sampling the CAV dynamics. This scenario relaxes the observability assumption as the information of every vehicle eventually reaches other vehicles between every two sampling time-instants because of the excessive rate of communication than the sampling rate. The single time-scale method, on the other hand, only performs one step of data-fusion and data-sharing between every two samples of system dynamics. This requires a specific network topology design to address the observability constraints. In this section, to allow for cheaper and less complicated communication/computation devices in vehicles, we propose a single time-scale observer to track the state of the CAV platoon. This requires communication/computation at the same rate of sampling. The proposed observer is composed of the following two parts:

\textit{(i) Consensus and averaging on predictions:}  
\begin{eqnarray}\label{eq_p}
	\widehat{\mb{x}}^i_{k|k-1} = \sum_{j\in\mathcal{N}(i)} w_{ij}A\widehat{\mb{x}}^j_{k-1|k-1},
\end{eqnarray}

\textit{(ii) Innovation update via sensor measurements:}     
\begin{eqnarray}\label{eq_m}
	\widehat{\mb{x}}^i_{k|k} =\widehat{\mb{x}}^i_{k|k-1} + K_k^i \sum_{j\in \mc{N}(i)}C_j^\top \left(\mb{y}^j_k-C_j\widehat{\mb{x}}^i_{k|k-1}\right),
\end{eqnarray}
where $\widehat{\mb{x}}^i_{k|k-1}$ (Resp. $\widehat{\mb{x}}^j_{k|k}$) denotes the state estimate at time-instant $k$ given all the information up-to time $k-1$ (Resp. up-to time $k$). The neighboring sets $\mathcal{N}(i)$ denote the set of neighbors of vehicle $i$. The block-diagonal gain matrix~${K}_k=\mbox{blockdiag}[K_k^i,\ldots,K_k^n]$ denotes the local observer gain at vehicles $1$ to $n$. Since the observer is distributed the gain matrix needs to be block-diagonal. Such a block-diagonal gain matrix can be computed via Linear Matrix Inequalities (LMI) designed based on iterative cone-complementarity optimization algorithms \cite{rami:97,usman_cdc:11}. Note that this observer only needs global observability of the CAV, and relaxes the assumption for local observability at every vehicle. In this regard, this work advances the papers by  \cite{das2015distributed,das2016consensus} in terms of observability assumption.

Denote the observer error at every vehicle by $\mb{e}_{k}^i := \mb{x}_{k|k} - \widehat{\mb{x}}^i_{k|k}$ and the column vector $\mb{e}_{k} = (\mb{e}_{k}^1,\dots,\mb{e}_{k}^n)^\top$ as the global error. The error dynamics of the proposed distributed observer \eqref{eq_p}-\eqref{eq_m} follows as,
\begin{align}
	\mb{e}_{k}^i &= \sum_{j\in \mathcal{N}(i)}w_{ij}A(\mb{x}_{k-1} - \widehat{\mb{x}}^j_{k-1|k-1})  \nonumber \\
	&- K_k^i \sum_{j\in \mathcal{N}(i)}C_j^\top C_j\sum_{j\in \mathcal{N}(i)}w_{ij}A(\mb{x}_{k-1}-\widehat{\mb{x}}^j_{k-1|k-1}) \nonumber \\ \label{eq_erri}
	 &+ \eta^i_k
\end{align}
and in compact form as,
\begin{align}\label{eq_err1}
	\mb{e}_{k} = (W\otimes A - K_k D_C (W\otimes A))\mb{e}_{k-1} +
	\eta_k,
\end{align}
with $\eta_k$ collecting the noise and input terms, operator $\otimes$ as the Kronecker matrix product, and
\begin{align}
	D_C := \left(
	\begin{array}{cccc}
		\sum_{j\in \mathcal{N}(i)}C_1^\top C_1\\
		&\ddots\\
		& &\sum_{j\in \mathcal{N}(i)}C_n^\top C_n\
	\end{array}
	\right)
\end{align} 
Following Kalman filtering theory \cite{kalman:61}, Eq.~\eqref{eq_err1} is steady-state Schur stable if the pair $(W \otimes A, D_C)$ is observable. In distributed estimation setup, this implies the notion of \textit{distributed observability}, i.e., the states of the vehicles need to be observable over the communication network $\mc{G}_W$ by sharing information locally in the neighborhood of vehicles. This observability model is less stringent than the local observability consideration in \cite{das2015distributed,das2016consensus} which requires system observability at every vehicle.
Following from the graph-theoretic results on network observability in \cite{tsipn}, one can satisfy distributed $(W\otimes A, D_C)$ observability by designing the matrix $W$ to be \textit{irreducible}, which implies that the communication graph $\mc{G}_W$ be strongly-connected. Note that this relaxes the network connectivity requirement in many existing works, e.g., the papers \cite{das2015distributed,das2016consensus} require more connectivity.

Given the error dynamics~\eqref{eq_err1}, the LMI for designing the block-diagonal observer gain is as follows: 
\begin{align} \label{eq_min}
	\begin{aligned}
		\displaystyle
		\min
		~~ &  \mathbf{trace}(XY) \\
		\text{s.t.}  ~~& X,Y\succ 0, ~ & K_k \mbox{~is~block-diagonal}.\\ ~ & \left( \begin{array}{cc} X&\widehat{A}^\top\\ \widehat{A}&Y\\ \end{array} \right) \succ 0,~& \left( \begin{array}{cc} X&I\\ I&Y\\ \end{array} \right) \succ 0,\\
	\end{aligned}
\end{align}
The stopping criteria of the above iterative LMI optimization is $\mathbf{trace}(Y_kX + X_kY)<2n^2 + \epsilon$ with a predefined small $\epsilon$ and $n^2$ as the size of the $\widehat{A}: = W\otimes A - K_k D_C (W\otimes A)$ matrix. We solve this iterative algorithm at the same time-scale $k$ of the CAV dynamics. The LMI~\eqref{eq_min} gives a $K_k$ such that $\rho (\widehat{A})<1$, where $\rho (\cdot)$ denotes the spectral radius of its argument. This implies that the discrete-time observer-based error dynamics~\eqref{eq_err1} is Schur stable.

\begin{figure} [hbpt!]
	\centering
	\includegraphics[width=2.25in]{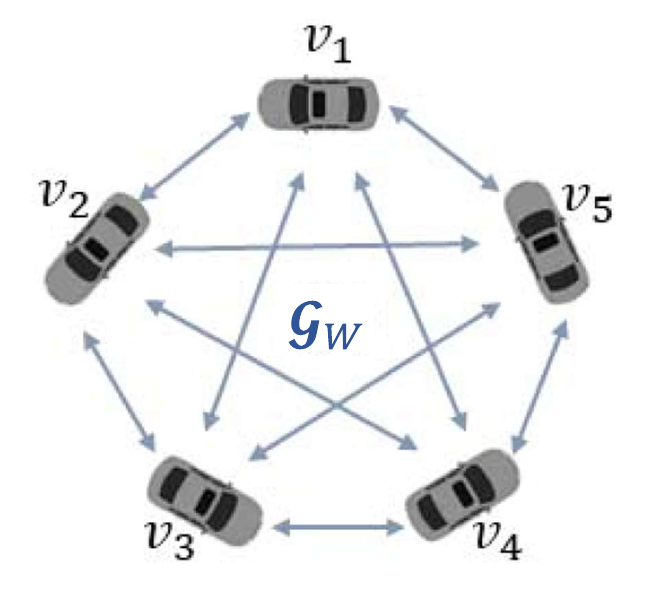}
	\caption{This figure shows a $3$-node/link-connected CAV network. This network preserves connectivity after the failure of up to $3$  nodes or $3$ links, and therefore, the distributed observer remains practical even in the failure of $3$ vehicle nodes or $3$ communication channels. }
	\label{fig_survivable}
\end{figure}

One can make the proposed distributed observer resilient to node and link failure by adding node/link redundancy in the communication network of vehicles \cite{ecc22_redund}. This is based on the notion of \textit{survivable network design} \cite{soni1999survivable} for $\kappa$-node-connected and $\kappa$-link-connected networks. Such robustified network designs are known to have at least $\kappa$ link-disjoint paths between any pair of vehicular nodes. In simpler terms, it means that to disconnect the network, one needs to remove at least $\kappa$ links or nodes \cite{cai1989minimum}. A $\kappa$-link/node-connected network is crucial for ensuring robustness against link/node failures or disconnections and enhances the reliability of information flow over the communication network of vehicles. Most existing algorithms for survivable network design are heuristic, for example, see \cite{grotschel1995design,diarrassouba2024optimization}. 
As an example in ITS literature, we refer interested readers to \cite{pirani2022impact} for understanding the impact of resilient network topology design on the CAV dynamics. An example survivable design via $\kappa$-node/link-connected CAV network (with $\kappa=3$) is shown in Fig.~\ref{fig_survivable}. This CAV network is resilient to failure of up to $3$ nodes or links, i.e., the network remains strongly-connected after removing up to $3$ failed communication channels or failed vehicles. Therefore, the distributed observer can track the state of the CAV even in case of such failures.

\section{Simulation Results} \label{sec_sim}
\begin{figure} [hbpt!]
	\centering
	\includegraphics[width=2.75in]{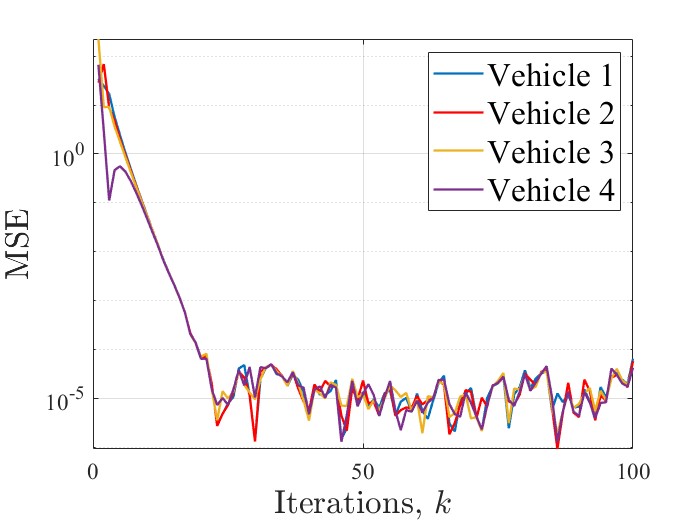}
	\caption{This figure shows the mean-square error at every vehicle under the proposed iterative distributed observer \eqref{eq_p}-\eqref{eq_m}. As it is clear from the figure the error dynamics is Schur stable with some steady-state residual due to Gaussian measurement noise. }
	\label{fig_mse0}
\end{figure}

For simulation, we consider the platoon setup in Fig.~\ref{fig_platoon} with $4$ vehicles and the represented $1$-link/node-connected communication network $\mc{G}_W$. The stochastic weight matrix associated with this setup is assigned via the simple averaging consensus protocol in \cite{olfati2007distributed}.  In system dynamic Eq.~\eqref{eq_xdot_d}, the sampling time of the dynamics is set $T=0.1$ and the driveline time-constant is $\tau = 0.5$. In the platooning setup by Eq.~\eqref{eq_di}, the static gap between vehicles is set as $D=5$ and the time-headway as $h = 1$. The noise in sensor measurement Eq.~\eqref{eq_yi} is considered as Gaussian $r \sim \mc{N}(0,0.04)$. The vehicle inputs follow the nearly-constant-acceleration model in \cite{icrom22_track}. Each vehicle applies the iterative distributed observer \eqref{eq_p}-\eqref{eq_m} to track the state of the CAV dynamics only by communicating information with its neighboring vehicles. The observer gain $K_k$ is defined via the LMI optimization \eqref{eq_min} iteratively over time-instant $k$. Using this gain, the resulting observer error dynamics has spectral radius $\rho (\widehat{A})=0.76<1$, which is Schur stable. The mean-square error (MSE) at every vehicle follows Eq.~\eqref{eq_erri} and is shown in Fig.~\ref{fig_mse0}. As clear from the figure, the error is stable in all vehicles, with some steady-state residual due to measurement noise.

\begin{figure} [hbpt!]
	\centering
	\includegraphics[width=2.75in]{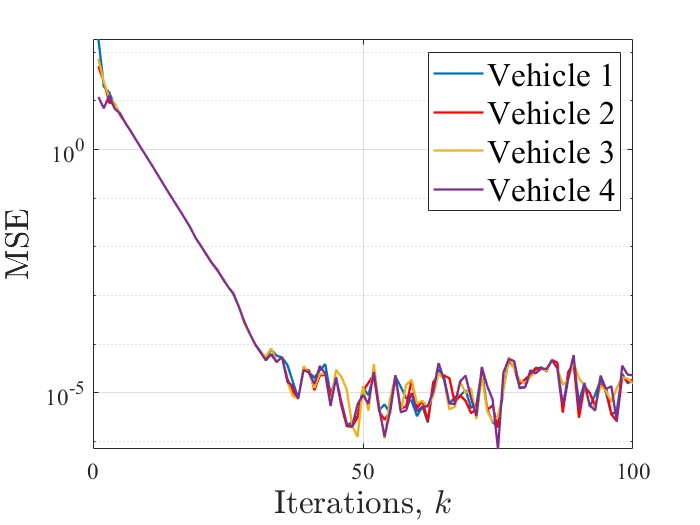}
	\caption{This figure shows the distributed observer error at every vehicle under the proposed iterative  \eqref{eq_p}-\eqref{eq_m} and reduced network connectivity, which is Schur stable. }
	\label{fig_mse1}
\end{figure}

For the next simulation, we consider the distributed estimation under link failure. Recall that the CAV network in Fig.~\ref{fig_platoon} is resilient to failure of $1$ arbitrary link, i.e., by removing any link the remaining CAV network is still strongly-connected. The strong-connectivity is a sufficient condition for the stability of the error dynamics under the proposed distributed observer setup. To show this by simulation, we remove the link from the $1$st vehicle to the $3$rd vehicle and check the MSE performance. The errors at all vehicles are shown in Fig.~\ref{fig_mse1} under the reduced CAV connectivity. The parameters are set the same as in the previous simulation. For this case, we have  $\rho (\widehat{A})=0.81<1$. The Schur stability of the error dynamics verifies the resilience of the proposed distributed observer to link failure.

\begin{figure} [hbpt!]
	\centering
	\includegraphics[width=2.75in]{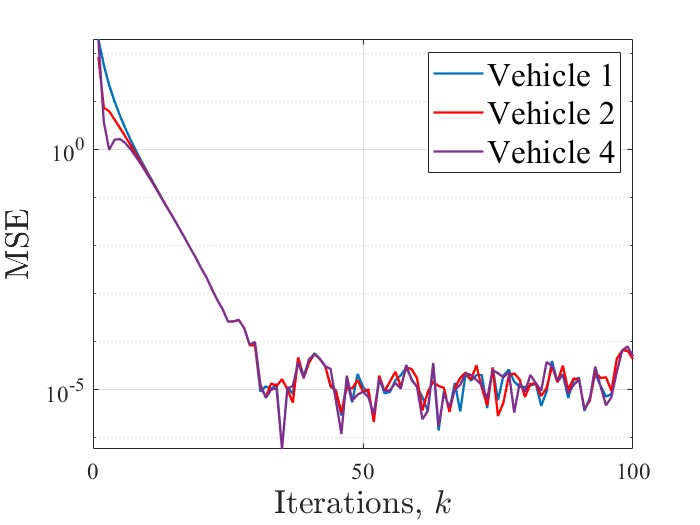}
	\caption{This figure shows the error at remaining $3$ vehicles under the proposed iterative distributed observer \eqref{eq_p}-\eqref{eq_m} after failure of the $3$rd vehicle. The Schur stability of the MSE shows the resilience of the proposed observer to node failure. }
	\label{fig_mse2}
\end{figure}

Next, we consider the distributed observer under node (vehicle) failure. Recall that the CAV network in Fig.~\ref{fig_platoon} remains strongly-connected after the failure of any $1$ vehicle. Removing the $3$rd vehicle and its associated links, we redo the MSE simulation for the other $3$ vehicles. Every remaining vehicle performs the iterative distributed observer to track the state of the CAV platoon excluding the $3$rd vehicle. The spectral radius of the distributed observer error dynamics for this simulation is $\rho(\widehat{A})=0.79<1$, which is Schur stable. The MSE performance is shown in Fig.~\ref{fig_mse2}. It is clear from the figure that the error decays along the iteration of the distributed observer. The steady-state residual is due to measurement noise in the CAV system.

\section{Conclusions and Future Directions} \label{sec_con}
This paper studies the tracking and estimation of vehicle platoons modelled as linear dynamical systems. The proposed iterative observer is distributed in the sense that each vehicle tracks the state of the CAV locally and with a decentralized technique, i.e., the estimation process is only based on local information exchange among neighboring vehicles. We clearly show that the error dynamics follows the Kronecker matrix product of the system matrix and adjacency matrix associated with the communication network of vehicles, and use graph-theoretic results to analyze its stability. In particular, we show that the strong-connectivity of the CAV network is sufficient for distributed observability and Schur stability of the error dynamics. In this direction, the redundant design of the network is proposed to make it resilient to node and link failure. LMI-based gain design is adopted to locally compute the observer gain matrix. The results significantly show that one can track the state of the CAV only using local information sharing over the network of vehicles. 

As future research direction, one may discuss the distributed observer design under network reliability concerns and packet drop  \cite{icrom_reliable}. Robustness to communication delays over the network is another important research direction. Integration with machine learning techniques as control input design and data-driven methods \cite{forgione2015data} to improve the tracking performance, particularly in scenarios with complex and uncertain dynamics, is another future research direction.

\bibliographystyle{IEEEbib}
\bibliography{bibliography}

\end{document}